\documentstyle[twocolumn,aps,psfig,newlfont]{revtex}
\begin{document}
\title{Sub-dekahertz ultraviolet spectroscopy of $^{199}$Hg$^+$ $^\dag$}
\author{R.~J.~Rafac, B.~C.~Young$^\ddag$, J.~A.~Beall, W.~M.~Itano, D.~J.~Wineland, and J.~C.~Bergquist}
\address{National Institute of Standards and Technology, Boulder, CO  80303 USA}
\maketitle
\begin{abstract}               
Using a laser that is frequency-locked to a Fabry-P\'{e}rot \'{e}talon of high finesse
and stability, we probe the
$5d^{10} 6s$ $^2$S$_{1/2}$(F=0)$\leftrightarrow$$5d^9 6s^2$ $^2$D$_{5/2}$(F=2) $\Delta$m$_F$=0
electric-quadrupole transition of a single laser-cooled $^{199}$Hg$^+$ ion stored in a cryogenic
radio-frequency ion trap.   We observe Fourier-transform limited linewidths as narrow as
$6.7$~Hz at 282~nm ($1.06\times 10^{15}$~Hz), yielding a line $Q\approx 1.6\times 10^{14}$.
We perform a preliminary measurement of the $5d^9 6s^2$ $^2$D$_{5/2}$
electric-quadrupole shift due to interaction with the static fields of the trap, and discuss 
the implications for future trapped-ion optical frequency standards.

\end{abstract}

\vspace{0.7cm}
Precision spectroscopy has held an enduring place in physics, particularly in the elucidation
of atomic structure and the measurement of fundamental constants, in the development of
accurate clocks, and for fundamental tests of physical laws.
Two ingredients of paramount importance are high accuracy, that is, the uncertainty in systematic
frequency shifts must be small, and high signal-to-noise ratio, since the desired measurement precision must
be reached in a practical length of time.  In this paper, we report
the measurement of an optical absorption line in a single laser-cooled $^{199}$Hg$^+$ ion
at a frequency $\nu_0=1.06\times10^{15}$~Hz (wavelength $\approx 282$~nm) for which a
linewidth $\Delta\nu =6.7$~Hz is observed, yielding the highest $Q\equiv \nu_0/\Delta\nu$
ever achieved for optical (or lower frequency) spectroscopy.  We also report a
preliminary measurement of the interaction of the upper state electric-quadrupole moment
with the static field gradients of the ion trap, which is expected to contribute the largest
uncertainty for a frequency standard based on this system.

In spectroscopy and for clocks, fluctuations in frequency measurement are usually expressed fractionally:
$\sigma_y(\tau) = \Delta\nu_{meas}(\tau)/\nu_0$, where $\tau$ is the total measurement time.  When the stability
is limited by quantum fluctuations in state detection,
$\sigma_y(\tau) =C(2\pi \nu_0)^{-1}(N\tau_{probe}\tau)^{-\frac{1}{2}}$, where N is the number of atoms,
$\tau_{probe}$ is the transition probe time (typically limited by the excited-state lifetime or the
stability of the local oscillator), and $C$ is a constant of order unity that depends on the method
of interrogation.  For many decades, the highest accuracies and the greatest stabilities have
been achieved by locking a microwave oscillator to a hyperfine transition in an atomic ground state
\cite{parisfount,jplhg,fisk,danahg,nistfount}.
Since the fractional instability $\sigma_y(\tau)$ is inversely proportional to the transition frequency,
greater stability can be attained using transitions at higher frequencies such as
those in the optical region of the electromagnetic spec-\linebreak
\begin{figure}
\vspace{-0.3in}
\begin{center}\leavevmode  
\psfig{figure=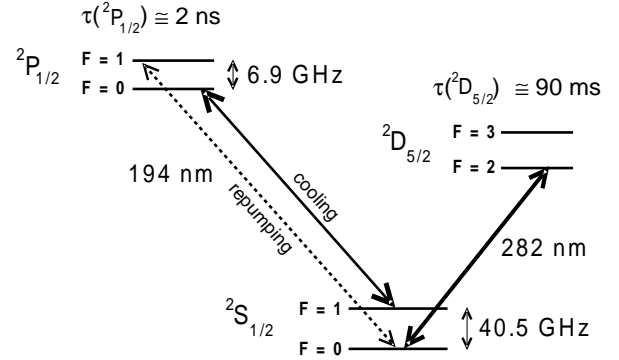,width=3.1in}
\end{center}
\caption[Calibration spectra]{Partial energy level diagram of $^{199}$Hg$^+$ with the transitions of interest
indicated.}
\label{hglevel}
\end{figure}
\noindent trum.  Only recently
have lasers of sufficient spectral purity become available to probe the narrow resonances
provided by transitions between long-lived atomic states
\cite{laserprl,salomon,sampas,acef,seel,picols,madej}.
The stable laser source of
Refs.~\cite{laserprl,picols} and the relative freedom from environmental perturbation afforded by ion trapping
enable the high resolution reported here.  Combined with novel, highly compact,
and accurate laser frequency measurement schemes \cite{udem,diddams}, a trapped-ion
optical frequency standard would appear to have significant advantages over present-day atomic
clocks.

A partial energy-level diagram of $^{199}$Hg$^+$ is shown in Fig.~\ref{hglevel}.
The 282~nm radiation used to drive the $^2$S$_{1/2}$$\leftrightarrow$$^2$D$_{5/2}$ transition is produced 
in a non-linear crystal as the second harmonic of a dye laser oscillating at 563~nm.
The frequency of the dye laser radiation is electronically servoed to match a longitudinal mode of a 
high-finesse Fabry-P\'{e}rot cavity that is temperature controlled and supported on an isolation
platform \cite{laserprl,picols}.
The stabilized laser light is sent through an optical fiber to the table holding the ion trap.
Unavoidable mechanical vibration of this fiber broadens the laser spectrum by nearly 1~kHz.
The fiber-induced phase-noise is sensed and removed using a method \cite{picols} similar to that
described in Ref.~\cite{jhall}.
Finally, the frequency of the 563~nm light is referenced to the
electric-quadrupole transition by first frequency-shifting in an acousto-optic modulator (AOM) and then frequency-doubling
in a single pass through a deuterated ammonium dihydrogen phosphate crystal.

The design of the linear cryogenic ion trap used in these measurements borrows heavily
from our previous work \cite{danahg,raizen,poitzsch}.  A single $^{199}$Hg atom from a thermal source is ionized by an
electron beam and trapped
in the harmonic pseudopotential formed by the rf and static potentials of a
linear quadrupole trap.  The trap operates in analogy to a quadrupole mass filter having its ends
``plugged'' with static fields.  The trap electrodes are constructed from Au-metallized
alumina tubes 530~$\mu$m in diameter.  The electrode axes are held parallel and coincident with
the vertices of a square of side $1.37$~mm by an alumina structure that facilitates electrical
connection and mounting in the vacuum chamber.

One pair of diagonally opposite electrodes is segmented by laser micro-machining prior to metallization
to permit application of the axially confining static potential. The remaining pair of electrodes is
driven by a cryogenic copper helical resonator coupled to a $8.6$~MHz signal source.  Under typical 
operating conditions, a single $^{199}$Hg$^+$ ion exhibits secular
motion at 1.45~MHz, 1.86~MHz, and 1.12~MHz in the $x$, $y$, and $z$ (axial) directions, respectively, 
as inferred from the vibrational sideband spectrum of the ion \cite{jimmoss}.  In addition, biasing electrodes
are mounted outside the trap rods to cancel any stray static electric fields that may be present.

Previous experiments using $^{199}$Hg were performed at room temperature and at a pressure of approximately
$10^{-7}$~Pa \cite{laserstab}.  Under those conditions, the background gas pressure was large enough that the ion would be
lost due to chemical reaction after only a few minutes.  To circumvent this, the ion trap is housed
in a liquid He vacuum Dewar like that described in \cite{poitzsch}.
Engineering particulars of the liquid He cryostat allow low-frequency ($< 100$~Hz) vibratory motion of the
trap structure relative to the optical table.  Uncorrected, these vibrations contribute $50-1000$~Hz of
Doppler broadening to the laser line.  We eliminate the majority of this broadening using an additional stage
of Doppler cancellation,
where the correction signal is derived from optical heterodyne detection of a motion-sensing beam reflected from a
mirror rigidly affixed to the trap \cite{laserstab}.  The resulting cancellation is not ideal, because
the sensing beam is steered by additional optical elements and its path
deviates slightly from overlap with the probe beam near the trap.  Measurements indicate that this optical
path difference can contribute as much as 2~Hz to the spectral width of the 282~nm probe
laser in the reference frame of the ion.

The ion is laser-cooled to near the 1.7~mK Doppler limit by driving the
$5d^{10} 6s$ $^2$S$_{1/2}$~(F=1)$\leftrightarrow$$5d^{10} 6p$ $^2$P$_{1/2}$~(F=0) cycling
transition at 194~nm (Fig.~\ref{hglevel}) \cite{jimmoss}.
Because of weak off-resonant pumping to the $^2$S$_{1/2}$~(F=0)
state, we employ a second 194~nm source phase-locked to the first
with a 47~GHz offset that returns the ion to the ground-state F=1 hyperfine level. We tolerate the
complication of hyperfine structure, since
only isotopes with nonzero nuclear spin can have first-order magnetic-field-insensitive transitions that
provide immunity from fluctuations of the ambient field.  This significantly relaxes the requirements
for control and/or shielding of environmental magnetic sources.  

We monitor the ion and deduce its electronic state using light scattered from the cooling transition.
Fluorescence at 194~nm is collected by a five-element uv-grade fused silica $f/1$ 
objective located inside the cryostat.  The scattered light is imaged outside the Dewar, spatially filtered
with a $75$~$\mu$m aperture, and relayed with a second lens to an imaging photomultiplier tube
having $\approx 5$\% quantum efficiency at 194~nm.  Transitions
to the $^2$D$_{5/2}$ state are detected using the technique of ``electron shelving,'' which infers the
presence of the atom in the metastable level through the absence of scattering from the strong laser-cooling
transition \cite{jimmoss,shelving}.   A metastable-state detection efficiency near unity can be achieved,
because the absorption of a single 282~nm photon suppresses scattering of many photons
from the 194~nm transition for a period determined by the lifetime of the $^2$D$_{5/2}$ state.
The radiation from the 194~nm and
282~nm sources is admitted to the trap sequentially using mechanical shutters and an AOM, which prevents
broadening of the quadrupole transition by the cooling radiation.
Typical count rates are $2000$~Hz for a single ion cycling at the half-power point of the cooling transition,
compared to only $20$~Hz combined laser scatter and photomultiplier thermal background when the ion is shelved
in the metastable level.

Spectra of the recoilless ``carrier'' component of the
$^2$S$_{1/2}$ (F=0)$\leftrightarrow$$^2$D$_{5/2}$ (F=2) $\Delta$m$_F$=0
transition were obtained for a range of probe times and laser intensities by laser cooling for $30$~ms,
preparing the ion in the F$=0$ ground state by blocking the repumping laser, and then interrogating the quadrupole
transition.  The spectra are built up from multiple bidirectional scans of the 282~nm probe laser frequency.
 Since the frequency drift of the probe
laser is not precisely compensated, nor is it constant, 
 we incorporate a locking step in between pairs of positive- and negative-going frequency sweeps about the
center of the quadrupole resonance.  In the locking sequence, we step the frequency of the probe laser
alternately to the approximate
half-maximum on either side of the quadrupole resonance, probe for a fixed time $\tau_{servo}$, and
then look for transitions to the $^2$D$_{5/2}$ level.  Typically, 48 measurements are made on each side of the
resonance during each lock cycle.  The asymmetry between the number of quantum jumps detected on the high- and
low-frequency sides of the resonance is used to correct the frequency of a synthesizer used to compensate for
cavity drift.  In this fashion, variations in the frequency of the 282~nm laser for times exceeding several seconds
are reduced.  Using the locking step alone with $\tau_{servo}=40$~ms, the error signal from the lock indicates
a fractional frequency instability $\sigma_y (\tau)=1.5\times 10^{-15}$ for times comparable to the length of the
combined servo and scanning cycle (15-60~s).  This is worse than the quantum projection-noise-limited stability
($\approx 5\times 10^{-16}$)
but is consistent with fluctuating quadratic Zeeman shifts arising from variations in the ambient magnetic field.
The trap was not magnetically shielded during the measurements reported here.

\begin{figure}
\begin{center}\leavevmode  
\psfig{figure=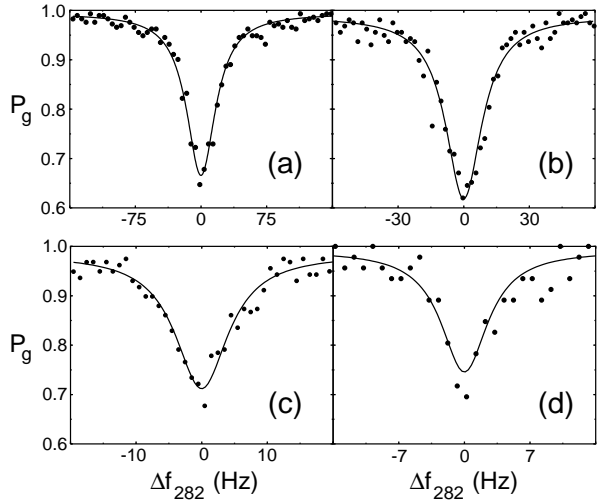,width=3.1in}
\end{center}
\caption[FT-limited]{Quantum-jump absorption spectra of the $^2$S$_{1/2}$ (F=0)$\leftrightarrow$$^2$D$_{5/2}$
(F=2) $\Delta$m$_F$=0 electric-quadrupole transition.  $\Delta f_{282}$ is the frequency of the 282~nm probe
laser detuning, and P$_g$ is the probability of finding the atom in the ground state.
The four plots correspond to excitation with 282~nm pulses of different lengths: (a) 20~ms (averaged over 292 sweeps),
(b) 40~ms (158 sweeps), (c) 80~ms (158 sweeps), and (d) 120~ms (46 sweeps).  The linewidths are consistent
with the Fourier-transform limit of the pulse at 40(2)~Hz, 20(1)~Hz, 10(1)~Hz, and 6(1)~Hz.}
\label{blockplot}
\end{figure}

Spectra are plotted in Fig.~\ref{blockplot} for a variety of probe times $\tau_{probe}=20-120$~ms, with
$\tau_{servo}=\tau_{probe}$.  The linewidths of all the spectra are transform limited by the
\noindent finite probe time, decreasing to $6.7$~Hz in the uv
at 120~ms, the longest time used.  The carrier transition amplitude is a function of the initial value
of the vibrational quantum number $n$ (see, for example, Eqn.~31 of Ref.~\cite{depth}).  Since for our trapping
parameters $\langle n \rangle \approx 35$ at the Doppler cooling limit, it is not possible to transfer the electron
to the $^2$D$_{5/2}$ state with unit probability.  The observed signals are in good agreement with the theoretical
expectation, and
the signal loss for $\tau_{probe}=120$~ms (Fig.~\ref{blockplot}(d)) is consistent
with applying the probe for a time that exceeds the natural lifetime of the $^2$D$_{5/2}$ state by $33\%$.
This result corresponds to a fractional frequency resolution of
$6.3\times 10^{-15}$, which we believe is the smallest reported for excitation with an optical or
microwave source.  The result is surpassed only by M\"{o}ssbauer spectroscopy in ZnO, where a fractional
linewidth of $2.5\times 10^{-15}$ was achieved using a nuclear source
of $93$~keV gamma rays \cite{mossbauer}, which is not practical for use in an atomic clock.
The measurements reported here illustrate
the potential of a single-ion optical frequency standard and confirm the phase stability
of our probe laser system, validating the results of the heterodyne comparison of the two
reference cavities \cite{laserprl,picols}.  An excellent review of the performance of similar systems can be found in
Ref.~\cite{madej}.

\begin{figure}
\vspace{-0.45in}
\begin{center}\leavevmode  
\psfig{figure=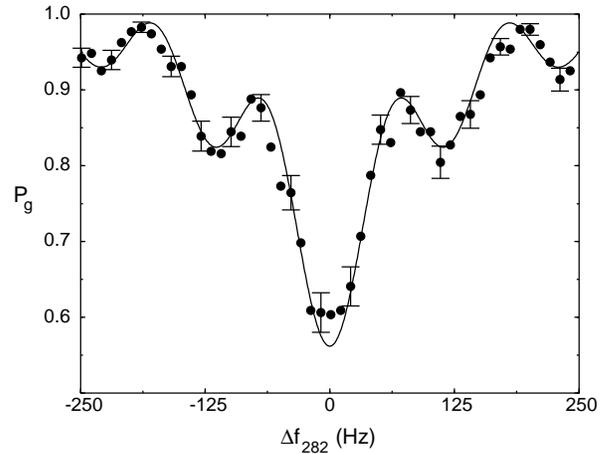,width=3.1in}
\end{center}
\caption[Rabi]{Quantum-jump absorption spectrum of the power-broadened
electric-quadrupole transition.  $\Delta f_{282}$ is the frequency of the 282~nm probe laser detuning, and P$_g$ is the probability
of finding the atom in the ground state.  The data are averaged over 348 frequency sweeps with $\tau_{probe}=10$~ms
and $\tau_{servo}=40$~ms.
The solid line is a least-squares fit to a Rabi line profile with pulse area 2.41(7)$\pi$ and
$\langle n \rangle$=73(9).  The quantum-projection-noise-limited uncertainty is indicated by representative error
bars plotted every third point.}
\label{rabiex}
\end{figure}

Figure \ref{rabiex} shows a typical power-broadened spectrum with consequent Rabi ``sidebands,'' using
$\tau_{probe}=10$~ms and $\tau_{servo}=40$~ms.  A least-squares fit to the 
data indicates that it is consistent with a pulse area of 2.41(7)$\pi$ (relative to $n\approx 0$) 
and a mean vibrational quantum num-\linebreak
\noindent ber $\langle n \rangle$=73(9).  This is twice the value of $\langle n \rangle$
expected at the Doppler cooling limit; the larger result is likely due to saturation of the cooling transition.

\begin{figure}
\begin{center}\leavevmode  
\psfig{figure=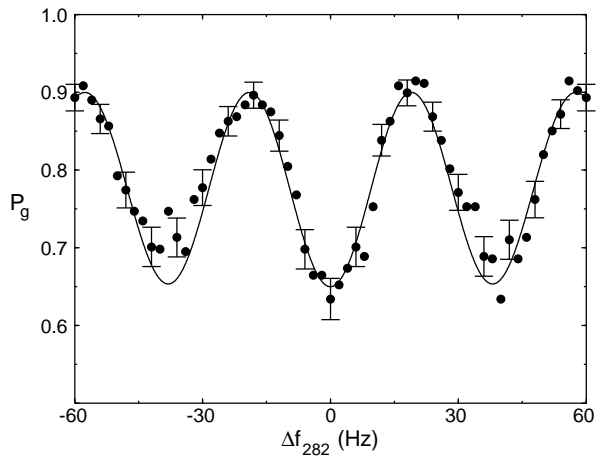,width=3.1in}
\end{center}
\caption[Ramsey]{The central three fringes obtained by time-domain Ramsey interrogation of the 
electric-quadrupole transition.  $\Delta f_{282}$ is the frequency of the 282~nm probe laser detuning, and P$_g$ is the probability
of finding the atom in the ground state.  The data are averaged over 328 frequency sweeps with $\tau_{servo}=40$~ms.
The solid line is the expected Ramsey signal for 5~ms pulses separated by 20~ms of free precession; the $\approx 10$\%
background arises primarily from suboptimal setting of the discriminator levels in our detection electronics.
The quantum-projection-noise-limited uncertainty is indicated by representative error
bars plotted every third point.}
\label{ramseyex}
\end{figure}

In Fig.~\ref{ramseyex} we plot the central three Ramsey fringes of a time-domain separated-oscillatory-fields
interrogation of the $^2$S$_{1/2}$$\leftrightarrow$$^2$D$_{5/2}$ transition with 5~ms pulses separated by a
20~ms free-precession interval.  In operation as a frequency standard, the Ramsey method offers $1.6 \times$
reduction in linewidth (and a corresponding increase in stability) for an equivalent measurement period.

It should be possible to reduce the uncertainties of all systematic shifts in this system including the second-order
Doppler (time-dilation) shift and static or dynamic Zeeman and Stark shifts to values approaching
$10^{-18}$.  The
electric-quadrupole shift of the $^2$D$_{5/2}$ F=2 (m$_F$=0) state arising from coupling with the
static potentials of the trap is expected to be the limiting systematic contribution in future
measurements of the absolute value of the $^2$S$_{1/2}$$\leftrightarrow$$^2$D$_{5/2}$ transition frequency,
because of the difficulty in determining with certainty the magnitude and configuration of the static
fields of the trap.  In principle, it is possible to eliminate this shift by measuring the
quadrupole transition frequencies for each of three mutually orthogonal orientations of a quantizing magnetic
field of constant magnitude; the mean value is then
the unperturbed $^2$S$_{1/2}$$\leftrightarrow$$^2$D$_{5/2}$ transition frequency.  We follow this procedure
for several values of the total magnetic induction and for all possible
directions in a suitable coordinate system (e.g.~[1 1 1], [1 -1 -1], [-1 -1 1], where [1 1 1]
is elevated $45^\circ$ from the trap axis and $45^\circ$ from the vertical plane).
Using the ion as a magnetometer, the magnitude of the field produced by three sets of coils is
\noindent calibrated by the measurement of the frequency of the F=$0 \rightarrow 2$ ($\Delta$m$_F=2$) magnetic
field-dependent electric-quadrupole transition.
For an ideal linear trap with secular frequencies $\omega_x=\omega_y$ and $\omega_z=2\pi$(1~MHz), we calculate 
a shift that varies between $+7.5$~Hz and $-15$~Hz depending on field orientation.  The estimate of the
atomic quadrupole moment is based on Hartree-Fock wavefunctions for the ground state of Hg$^+$, which yield
the matrix element $\langle 5d |r^2| 5d \rangle=6.61\times 10^{-21}$~m$^2$ \cite{mclean}.
In our apparatus we observe shifts departing from the mean frequency by
$+12(7)$~Hz, $+9(13)$~Hz, and $-27(7)$~Hz for the three field orientations at 282 nm, in reasonable agreement
with the simple model.
The quoted uncertainties arise from imprecision in our ability to set the
absolute magnitude and direction of the magnetic induction.
Thus the fractional frequency uncertainty is of order $10^{-14}$, and we might reduce this below
$10^{-15}$ in the present apparatus with straightforward improvements to the magnetic
shielding and control.  Uncertainties approaching $10^{-18}$ should be obtainable through
stringent control of magnetic fields in combination with
a spherical rf quadrupole trap geometry that does not rely on static potentials for confinement.
Greater resolution and accuracy might be more readily achieved using a different type of transition,
e.g., the weak hyperfine-induced electric-dipole transitions like those between the
low-lying $^1$S$_0$ and $^3$P$_0$ states of the singly-ionized species of the Group IIIA elements
of the periodic table \cite{madej,dehmeltIII,peik}, particularly in cases where first-order magnetic-field independent
transitions are available (albeit at nonzero field) \cite{davewayne}.
   
This research was partially supported by the Office of Naval Research and through a Cooperative
Research and Development agreement with Timing Solutions, Corp., Boulder, Colorado.  We would like to thank
C.\ Oates, J.\ Ye, and D.\ Sullivan for their careful reading of this manuscript.

\vspace{0.25cm}
\noindent $^\dag$Work of the U.S.~Government, not subject to U.S.~copyright.

\noindent $\ddag$Present address:  Jet Propulsion Laboratory, Pasadena, CA.


%
%

\newpage

\newpage

\end{document}